\documentclass[aps,floatfix,twocolumn,showpacs,amsmath,prl]{revtex4}
\usepackage{graphicx,bm}
\usepackage[T1]{fontenc}
\usepackage{ifthen}
\usepackage{psfrag}
\usepackage{color}

\newcommand{\REM}[1]{\ifthenelse{0=1}{#1}{}}
\begin{document}
\title{Solitons as the early stage of quasicondensate formation during evaporative cooling}

\author{E. Witkowska} 
\affiliation{Institute of Physics, Polish Academy of Sciences, Al. Lotnik\'ow
32/46, 02-668 Warsaw, Poland}

\author{P. Deuar} 
\affiliation{Institute of Physics, Polish Academy of Sciences, Al. Lotnik\'ow
32/46, 02-668 Warsaw, Poland}

\author{M. Gajda} 
\affiliation{Institute of Physics, Polish Academy of Sciences, Al. Lotnik\'ow
32/46, 02-668 Warsaw, Poland}

\author{K. Rz\k{a}\.zewski} 
\affiliation{Center for Theoretical Physics, Polish Academy of Sciences, Al. Lotnik\'ow
32/46, 02-668 Warsaw, Poland}

\date{\today}

\begin{abstract}
We calculate the evaporative cooling dynamics of trapped one-dimensional Bose-Einstein condensates for parameters leading to 
a range of condensates and quasicondensates in the final equilibrium state. 
We confirm that solitons are created during the evaporation process, but always eventually dissipate during thermalisation. 
The distance between solitons at the end of the evaporation ramp matches the coherence length in the final thermal state. 
Calculations were made using the classical fields method. They bridge the gap between the phase defect picture of the Kibble-Zurek mechanism and the long-wavelength phase fluctuations in the thermal state. 
\end{abstract}


\maketitle

The quasi-one-dimensional (1D) Bose gas in elongated clouds of neutral ultra-cold atoms\cite{1st-experiments-which-produced-1D-condensates,Petrov00} differs markedly from the Bose-Einstein condensate (BEC) in three-dimensional geometries. One of the most remarkable features is the presence of 
two characteristic temperatures when the trapped gas is cooled\cite{Petrov00}. Below $T_c$, the lowest mode becomes appreciably occupied\cite{Ketterle96} but the 
phase coherence length $l_{\phi}\propto1/T$ is smaller than the size of the system. This is called a quasicondensate. Below a second temperature $T_{\phi}$,  $l_{\phi}$ grows to the size of the cloud, and the state is a true BEC. 
The phase coherence in these states have been extensively studied both experimentally\cite{Richard03,phase-fluctuations-in-1D-experiments} and theoretically\cite{Ketterle96,Petrov00,Petrov01,Kadio05,Zurek09,Damski10}. 

In thermal equilibrium, the variance in phase and one-body density matrix 
have been calculated\cite{Petrov00,Petrov01,Dettmer01,Kadio05}. Their short-range behaviour gives the phase correlation length $l_{\phi}$ 
 near the center of the trap:
\begin{equation}\label{rho_1}
\rho_{1}(z,z') =\langle\hat{\Psi}^{\dagger}(z)\hat{\Psi}(z')\rangle\approx \rho_{1}(0,0)e^{-|z-z'|/l_{\phi}}.
\end{equation}
In equilibrium, the phase in a single experimental realisation varies smoothly over length scales $l_{\phi}$\cite{Dettmer01}. 

On the other hand, phase fluctuations have also been predicted from the Kibble-Zurek mechanism\cite{Kibble-Zurek-mechanism} (KZM) after the onset of condensation, when the system is far from equilibrium. These fluctuations are seemingly different in nature than those discussed above. 
During evaporative cooling, phase defects in the form of grey solitons appear when crossing the characteristic temperature $T_c$. They are born when local condensation occurs faster than distant regions can communicate to agree on a common phase. When the expanding initial phase domains meet, soliton defects form on the interfaces between them. 
Therefore, during the formation of a condensate the phase experiences sudden jumps at the temporal position of every soliton, and phase domains appear between them of a size equal to the separation between neighbouring solitons.  A natural question arises whether these pre-formed domains are somehow related to the phase fluctuations in equilibrium.  The aim of this paper is to show how the phase fluctuations in these two cases are connected. 

The number of solitons while crossing $T_c$ has been predicted as a function of the quench rate\cite{Zurek09}.  
A calculation where chemical potential was quenched at $T=0$ demonstrated the KZM for a uniform gas\cite{Damski10}. 
Here we show that the Kibble-Zurek scaling also applies for a realistic model of evaporative cooling in a trap. 
We have simulated the non-equilibrium evaporative cooling dynamics 
to the stationary thermal state using the classical fields method (CFM)\cite{cfm}. 
The observations imply that the solitons are indeed the early stage of development of the final phase fluctuations at equilibrium. 
Let us now proceed to the details.

\begin{figure}
\centerline{\includegraphics[width=5cm]{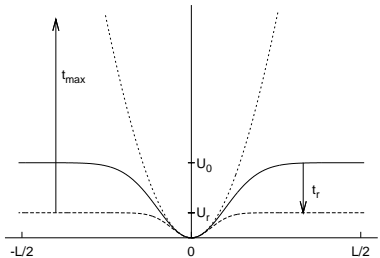}}
\vspace*{-0.3cm}
\caption{
\label{U}
The evaporative cooling potential $V(z,t)$. Solid line: initially ($t=0$), dashed line: at the end of the ramp ($t=t_r$), dotted line: during thermal equilibration ($t\gg t_r$). 
}
\end{figure}

We consider a single-species Bose gas in a trap. We use harmonic oscillator units with frequency $\omega_z$. 
The initial state of $N=10^4$ atoms is generated in a canonical ensemble at a temperature of $k_BT = 360\hbar\omega_z$\cite{MC}, well above $T_c$ in the final state.
The description of the system in the CFM is in terms of an ensemble of classical field amplitudes $\psi(z)$ evolving under a generalised Gross-Pitaevskii mean field evolution equation from thermally randomised initial conditions. This is equivalent to a truncated Wigner description\cite{truncWigner}, but with quantum fluctuations omitted. 
It is a good approximation for the 1D gas in our regime, where even the quasicondensate phase fluctuations are dominated by the thermal component\cite{Petrov00}. 
The evolution in time is given by 
\begin{equation}
i \partial_t \psi(z, t) = \left[ H(z,t) -i \Gamma(z,t)\right] \psi(z, t) \, ,
\end{equation}
with the Hamiltonian 
\begin{equation}
H(z,t) = -\frac{1}{2}\frac{\partial^2}{\partial z^2} + V(z,t) +g_{1D}|\psi(z,t)|^2
\end{equation}
including kinetic energy, contact interactions with strength $g_{1D}=0.31$ (corresponding to $^{87}$Rb atoms in a $10\times1000\times1000$ Hz trap), and the time-dependent external potential (shown in Fig.~\ref{U})
\begin{subequations}
\begin{eqnarray}
V(z,t) &=& U(t) \left[ 1 - e^{-\frac{z^2}{2U(t)}} \right]\,\\
U(t)&=&\left\{ \begin{array}{ll}
U_0 + (U_{r}-U_0) \frac{t}{t_{r}}, & t\le t_{r} \\
U_{r} + (U_{\rm max}-U_{r})\,\frac{t - t_r}{t_{\rm max}-t_r}, & t >t_r 
\end{array}\right .
\end{eqnarray}\end{subequations}
The first stage ($t<t_r$) is the evaporative cooling ramp. $V(z)$ is a Gaussian dip of constant trap frequency ($\omega_z=1$ in the units chosen) near $z=0$, and standard deviation $\sqrt{U(t)}$. The depth of the dip decreases linearly  from $U_0=100$ to $U_r=U_0/3$. The ramp time $t_r$ is varied to obtain different final states. 

After the ramp, ($t_r<t<t_{\rm max}$\,) the evaporation is stopped and the gas is allowed to thermalise for a longer period to $t_{\rm max}=t_r+1000$ in a much deeper potential (rising to a depth $U_{\rm max}=10U_0$) that becomes effectively harmonic in the region occupied by the gas cloud.  
It is also necessary to include losses in the wings of the potential: 
$\Gamma(z,t) = \Gamma_{\infty} \left[ V(z,t)/U(t)\right]^{\gamma}$
with $\Gamma_{\infty}=10$, $\gamma=50$. This loss acts as a high-energy knife in the region beyond about two ($t$-dependent) standard deviations of the Gaussian dip. This is necessary to realistically  model the experimental properties of the trap by preventing once evaporated atoms from returning back. 
A lattice of 1024 points on a length $L=120$ is used. 
The number of trapped atoms we obtain for $t\ge t_r$  is always around $N\approx1300$. 
The widths and shapes of the clouds 
agree very well with the radius $z_{TF}=\sqrt{2\mu}$ 
and chemical potential $\mu=[9(g_{1D}N)^2/32]^{(1/3)}$ predicted by the Thomas-Fermi approximation. 

\begin{figure}
\begin{picture}(8.6,11.5)
\put(0,8.5){\includegraphics[width=4cm]{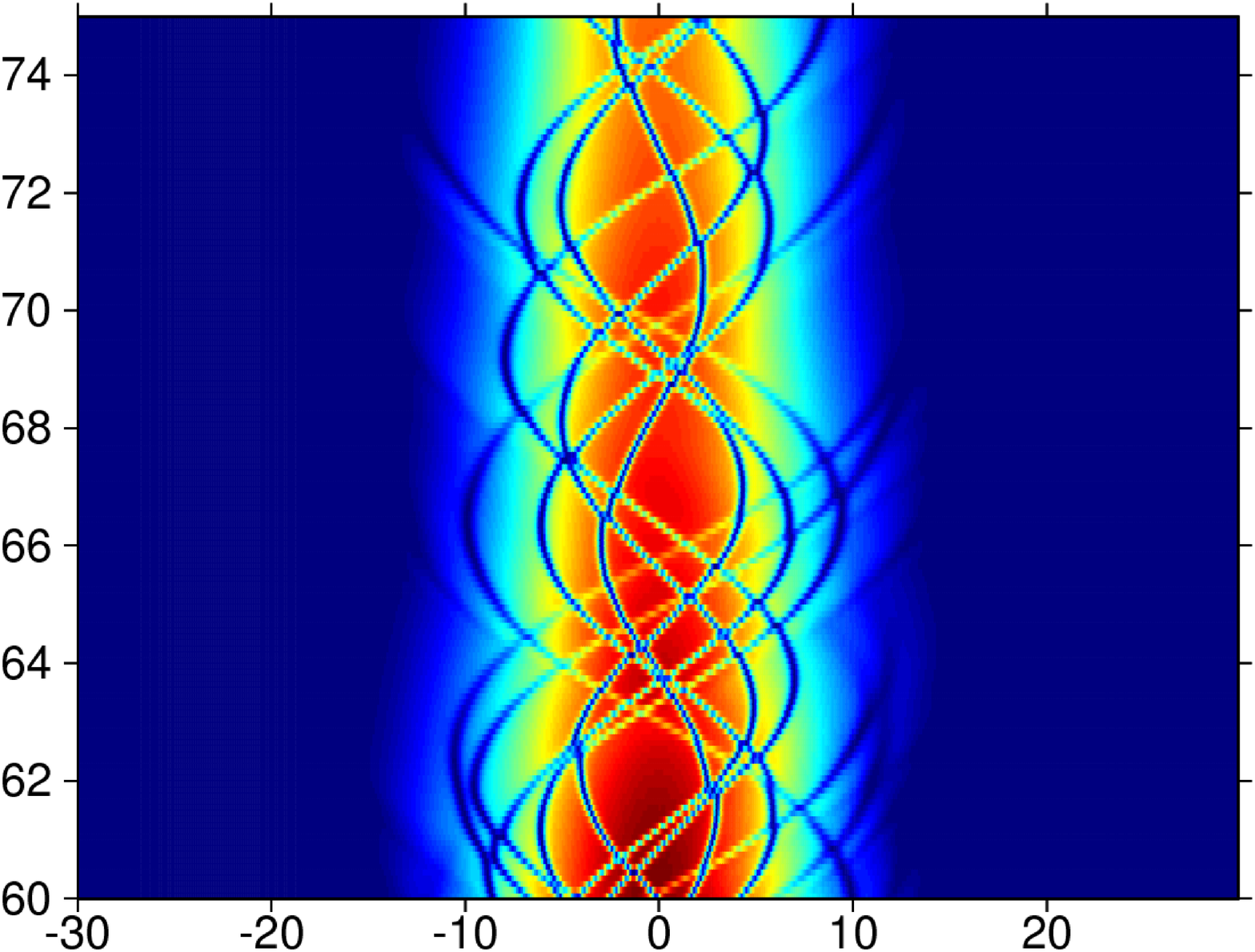}}
\put(4.4,8.5){\includegraphics[width=4cm]{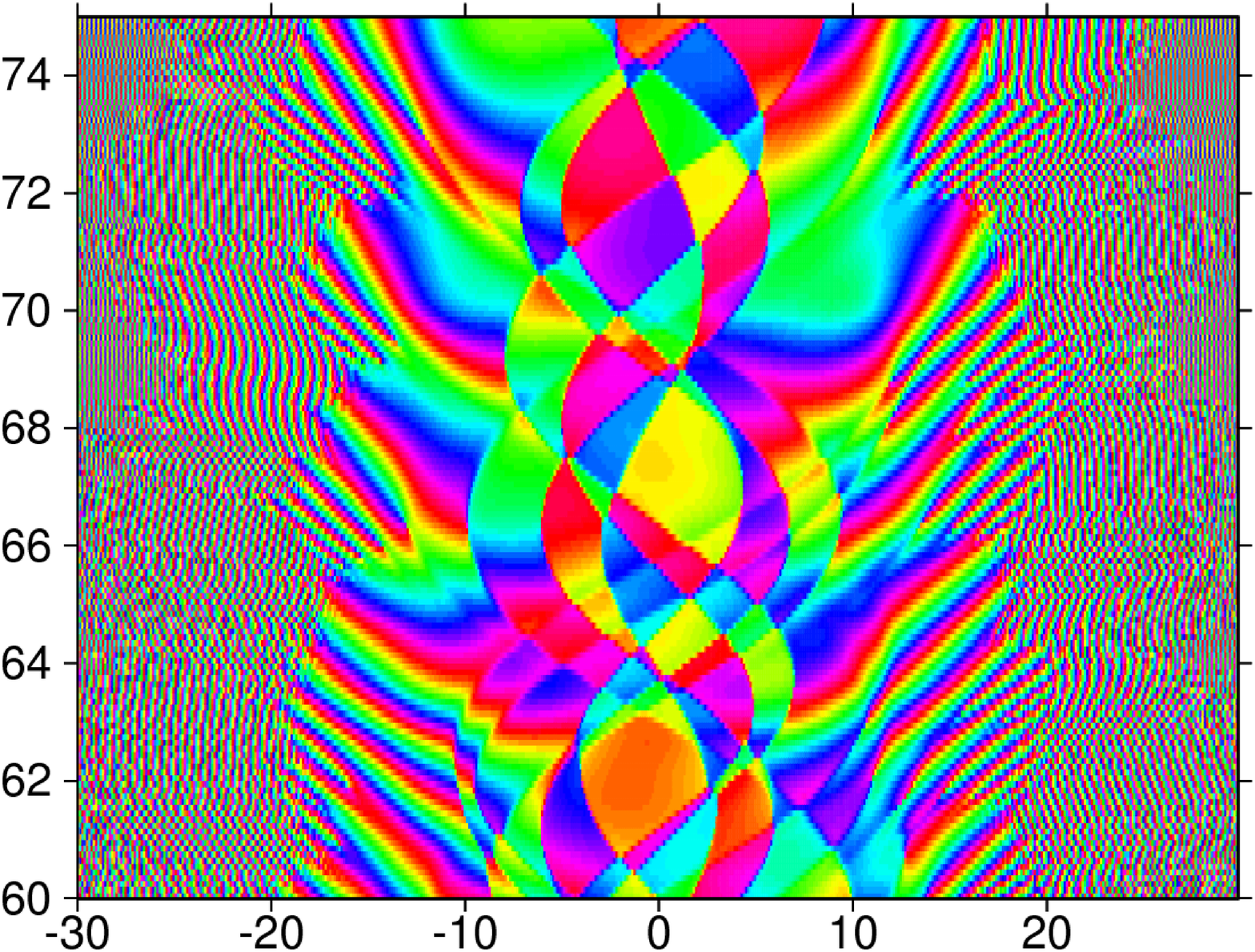}}
\put(0.3,11.15){\textcolor[rgb]{1,1,1}{\bf(a)}}
\put(4.8,11.15){\textcolor[rgb]{1,1,1}{\bf(b)}}
\put(-0.25,10.0){\rotatebox{90}{$t$}}
\put(3.1,10.0){\textcolor[rgb]{1,1,1}{$\scriptstyle t_r=75$}}

\put(0.8,3.2){\includegraphics[width=6.9cm]{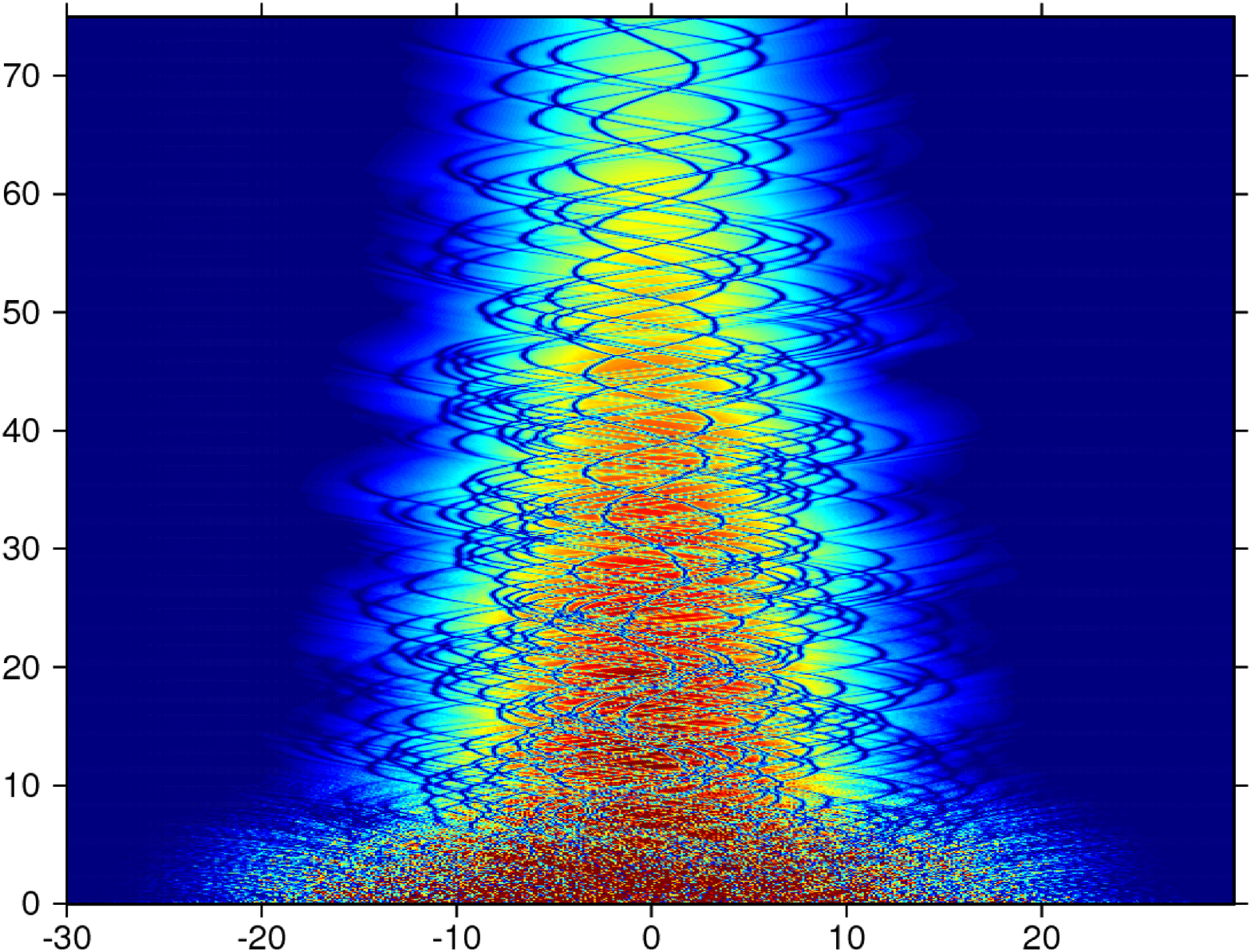}}
\put(1.25,8.0){\textcolor[rgb]{1,1,1}{\bf(c)}}
\put(4.3,3.0){$z$}
\put(0.5,5.8){\rotatebox{90}{$t$}}
\put(6.3,5.8){\textcolor[rgb]{1,1,1}{$t_r=75$}}

\put(0,0){\includegraphics[width=4cm]{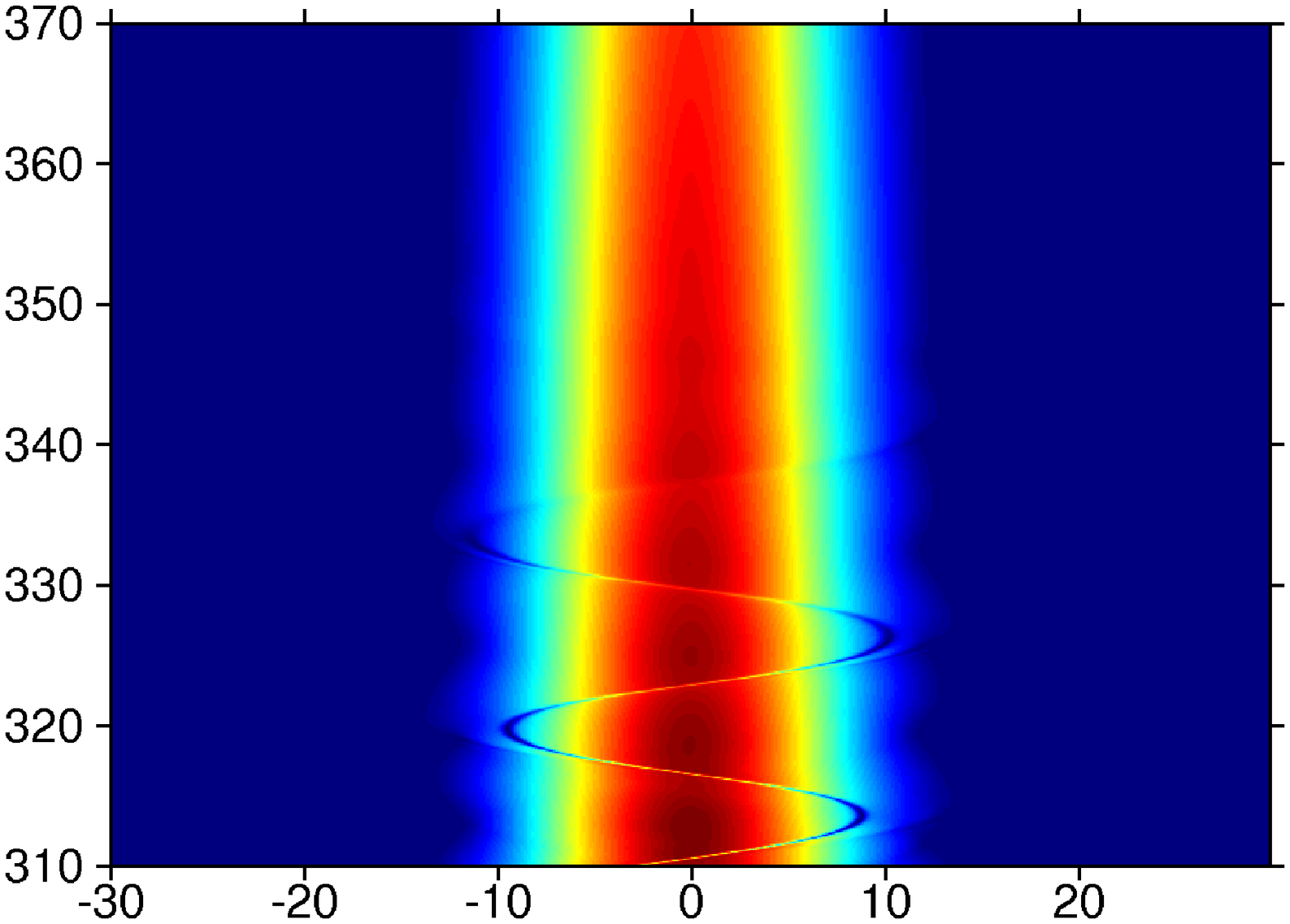}}
\put(4.4,0){\includegraphics[width=4cm]{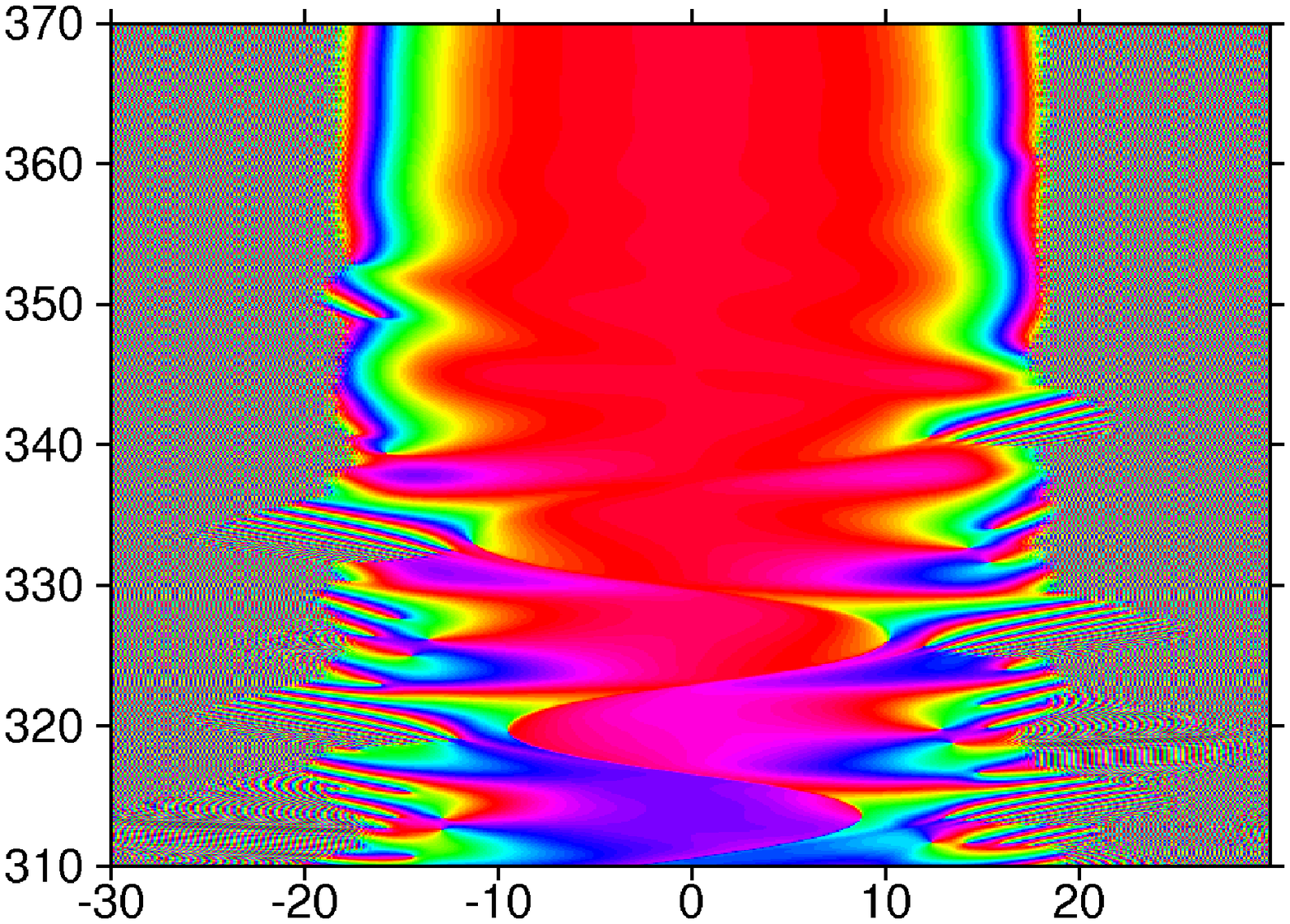}}
\put(0.35,2.5){\textcolor[rgb]{1,1,1}{\bf(d)}}
\put(4.8,2.5){\textcolor[rgb]{1,1,1}{\bf(e)}}
\put(-0.25,1.4){\rotatebox{90}{$t$}}
\put(3.0,1.4){\textcolor[rgb]{1,1,1}{$\scriptstyle t_r=400$}}
\end{picture}
\vspace*{-0.3cm}
\caption{
\label{tr75}
Density (a,c,d), and phase modulo $2\pi$ (b,e) for a single realisation during the cooling ramp. In (a-c) $t_r=75$ leading to a quasicondensate as $t\to\infty$, 
in (d-e) $t_r=400$ leading to  a BEC. 
Top: at the end of evaporative cooling. 
Center: The entire ramp time. 
Bottom: the loss of the last soliton, leading to BEC formation. 
}
\end{figure}

Figure~\ref{tr75} shows the time-evolution of a single realisation of the experiment in the quasicondensate ($t_r=75$) and BEC ($t_r=400$) regimes. Animations for $t_r=250$ are in\cite{Supp}.
We see the emergence of a great number of defects on the healing-length scale in the early part of the cooling phase (e.g. around $t=6-10$ in Fig~\ref{tr75}(c)). The density dips in Figs.~\ref{tr75}(a,c,d) 
display all the characteristics of grey solitons: passing through each other, turn-around near the edge of the cloud when the central density reaches zero, motion much slower than the speed of sound, phase jumps, and width in agreement with the healing length $\xi = 1/\sqrt{g_{1D}(\rho_0(x)-\rho_{\rm min})}$. Here $\rho_0(x)$ is the local background density, $\rho_{\rm min}$ the minimum density. The density profile of the dips is a good match to the solitonic solution ($\rho_{\rm min}+(\rho_0-\rho_{\rm min})\tanh^2\left[(z-z_0)/\xi\right]$, where the dip is centered at $z_0$. Note the broadening of the solitons when they approach the edges of the cloud in Fig.~\ref{tr75}(a,d). Very clear phase domains are seen between the solitons in Figs.~\ref{tr75}(b), and account for the majority of the spatial phase variation.

\begin{figure}
\centerline{\includegraphics[width=8cm]{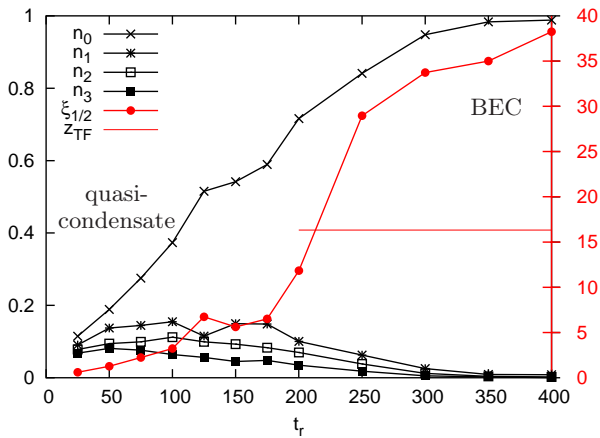}
\hspace*{-7.6cm}\raisebox{3cm}{\parbox{2cm}{quasi-\\condensate}}\hspace*{7.6cm}
\hspace*{-4cm}\raisebox{4.3cm}{BEC}\hspace*{4cm}\hspace*{-3cm}
}\vspace*{-0.3cm}
\caption{
\label{set}
Fraction of atoms in the four lowest energy modes $n_{0,1,2,3}$ (black), and phase domain width $l_{1/2}$ (red) at the end of the evaporative cooling ($t=t_r$) as a function of the ramp time $t_r$. The horizontal line (red) shows the Thomas-Fermi width ($2z_{TF}=2\sqrt{2\mu}$) achieved as $t\to t_{\rm max})$.
The ensemble consisted of 100 independent realisations. 
$l_{1/2}>2z_{TF}$ occurs when phase coherence extends into the tails of the cloud. 
}
\end{figure}

As cooling progresses, many of the solitons are lost at the edges, and the size of phase domains grows. We observe that if the final state achieved is a quasicondensate (see Fig.~\ref{set}), some solitons remain at $t_r$, while in the BEC regime (large $t_r$) all are already gone by this stage. 
Fig.~\ref{set} quantifies the crossover from quasicondensate to BEC as a function of ramp time $t_r$.

The number of condensed atoms $N_0(t)$ is calculated from the maximal eigenvalue of the one-particle density matrix $\rho_1(z,z')$. Here, this is calculated as $\langle \psi^*(z) \psi(z')\rangle_{\rm s}$, with $\langle\cdot\rangle_{\rm s}$ denoting averaging over realisations within the initial canonical ensemble. The condensate fraction $n_0=N_0/N$, is shown as the '$\times$' data in Fig.~\ref{set}. 
Phase correlations are characterised using the correlation function 
$g^{(1)}(z, z')=\rho_1(z,z')/\sqrt{\rho_1(z,z)\rho_1(z',z')}$. 
We use the width of the correlation function symmetric around zero to measure the phase domain size, using:
$g^{(1)}(-\frac{1}{2}l_{1/2},\frac{1}{2}l_{1/2}) = \frac{1}{2} \approx e^{-l_{1/2}/l_{\phi}}$. Hence, comparing to (\ref{rho_1}), $l_{\phi}=l_{1/2}/\log 2$.
	
Fig.~\ref{set} shows the transition between quasicondensate and BEC behaviour\cite{Petrov00}. In the quasicondensate, the phase domain width is much smaller than the cloud, and a sizable part, or even a majority of the atoms are not condensed into the lowest energy mode. In the condensate, a single phase domain covers the entire cloud, and the condensate accounts for the vast majority of the atoms. Notably, the transition in phase correlation length $l_{1/2}=l_{\phi}\log2$ around $t_r=200$ is much sharper than in condensate fraction $n_0$. 

As the state is non-thermal during the evolution, assigning a time-dependent temperature is moot, and we use the condensate fraction $n_0$ (always a well defined quantity, unlike $T$) as the analogous parameter. For interpretation, several approximate results will be useful\cite{Ketterle96}: $n_0\approx 1- T/T_c$, when $n_0$ is not too close to zero, and where the characteristic temperature $T_c$ is given by $N=T_c\log(2T_c)$. For $N=1300$, this is $T_c=214$. 
Hence, the temperature appearing in estimates of the coherence length\cite{Petrov00,Petrov01} can be replaced by 
\begin{equation}\label{Tn0}
T\to T_c(1-n_0).
\end{equation}
The location of the BEC/quasicondensate transition in Fig.~\ref{set} can be compared using (\ref{Tn0}) with the  prediction $T_{\phi}=N/\mu$\cite{Petrov00}. The latter leads to $n_0(T_{\phi})=1-N/(\mu T_c)\approx0.84$, which compares favourably with the data.

\begin{figure}
\centerline{\includegraphics[width=8cm]{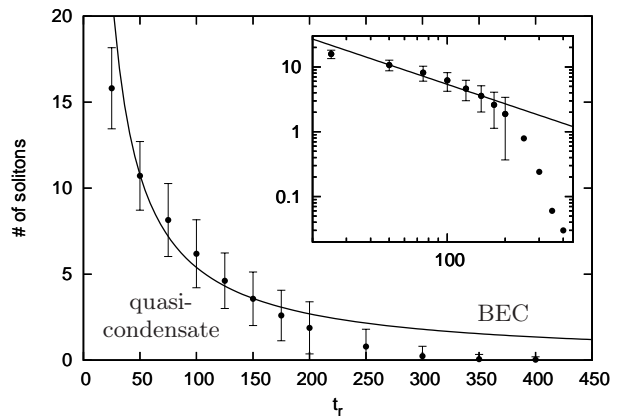}
\hspace*{-7.1cm}\raisebox{1.3cm}{\parbox{2cm}{quasi-\\condensate}}\hspace*{7.1cm}
\hspace*{-4cm}\raisebox{1.3cm}{BEC}\hspace*{4cm}\hspace*{-3cm}
}\vspace*{-0.3cm}
\caption{
\label{sol}
The mean number of solitons at $t_r$ (circles; error bars show standard deviation of the ensemble values). 
The solid line shows the scaling $N \propto 1/t_r$ approximately expected for mean-field systems from \cite{Zurek09}, 
fitted to the data in the quasicondensate regime ($t_r\le200$). The prefactor is  $\approx$540. The inset shows a logarithmic scale.
}
\end{figure}

To verify that the Kibble-Zurek mechanism (KZM) is at work here, we wish to compare the scaling of the number of solitons (predicted in \cite{Zurek09}) with the rate that the characteristic temperature $T_c$ is crossed. 
We count solitons by fitting the density in the solitonic solution in the same manner as in \cite{Damski10}, using a local background density. 
To make the comparison, we should also calculate the quench time of the relative temperature $\tau_Q$ in which the KZM is expressed\cite{Zurek09}. The prediction is 
$N_{\rm soliton} \propto \tau_Q^{-(1+2\nu)/(1+\nu z_{\tau})}$
where $\frac{dT}{dt} = -T_c/\tau_Q$, and $\nu$ and $\nu z_{\tau}$ are critical exponents for healing length and relaxation time at $T=T_c$, respectively. Here, because the system is certainly not in thermal equilibrium during the evaporative cooling phase, neither a value for $T_c$ nor $\frac{dT}{dt}$ is forthcoming. 
The most reasonable straightforward estimation is that the quench time is proportional to the ramp time: $\tau_Q\propto t_r$. This leads us to use $t_r$ as 
the ``quench time'' parameter in Fig.~\ref{sol} where the scaling is examined.

We see that the KZM power law scaling holds quite well in the quasicondensate regime for power law exponent unity in $1/t_r$, which is the value expected for mean field\cite{Zurek09}. 
As one might expect, the scaling law breaks down once the true condensate is reached for $t_r\gtrsim 200$, since the correlation length reaches or exceeds the size of the system
in the BEC regime. This suppresses the slowing down near $T_c$ that is an assumption of the KZM. 

In contrast to the elongated 3D case of \cite{Zurek09}, no condensation front is apparent here. This may be an effect of the slow internal thermalisation of 1D gases. 
We conjecture that a front might still occur in 1D if the local temperature were set externally by contact with an external reservoir, as e.g. in buffer gas cooling, but is unlikely in the type of scenario simulated here.

\begin{figure}
\begin{picture}(8,8.5)
\put(0,4.6){\includegraphics[width=8cm]{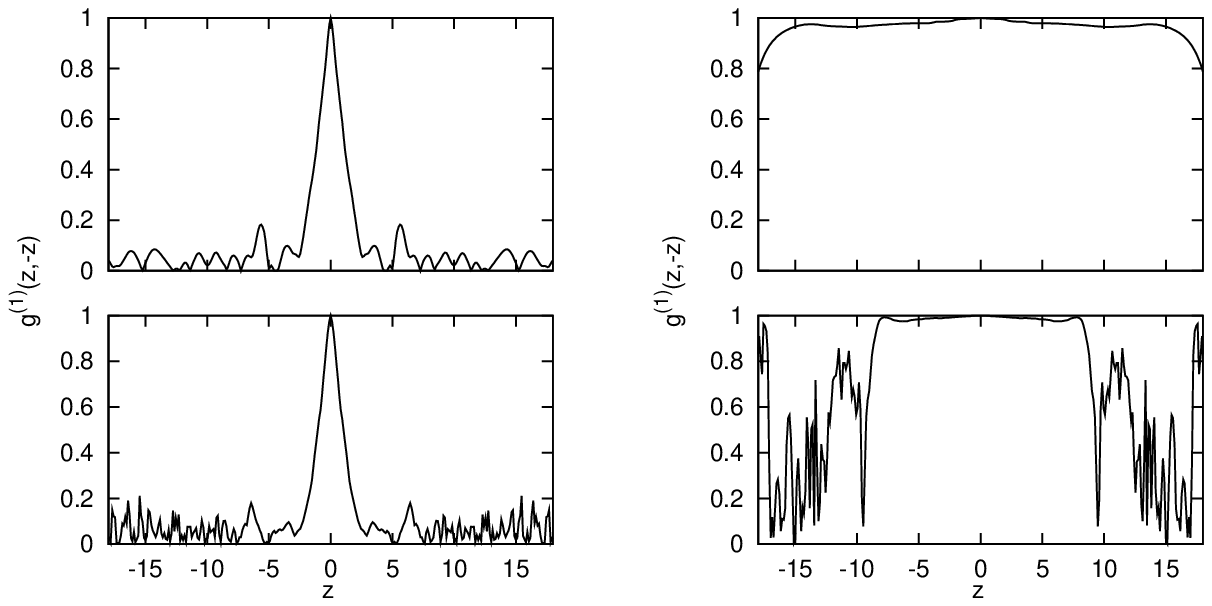}}
\put(0.75,8){\bf(a)}
\put(4.95,8){\bf(b)}
\put(0.75,6.1){\bf(c)}
\put(4.95,6.1){\bf(d)}
\put(0.45,8.55){quasicondensate: $t_r=75$}
\put(5.4,8.55){BEC: $t_r=400$}
\put(2.6,8){$t=t_r$}
\put(6.8,8){$t=t_r$}
\put(2.3,6.1){$t=t_{\rm max}$}
\put(5.8,6.1){$t=t_{\rm max}$}
\put(0.8,-0.1){\includegraphics[width=6.5cm]{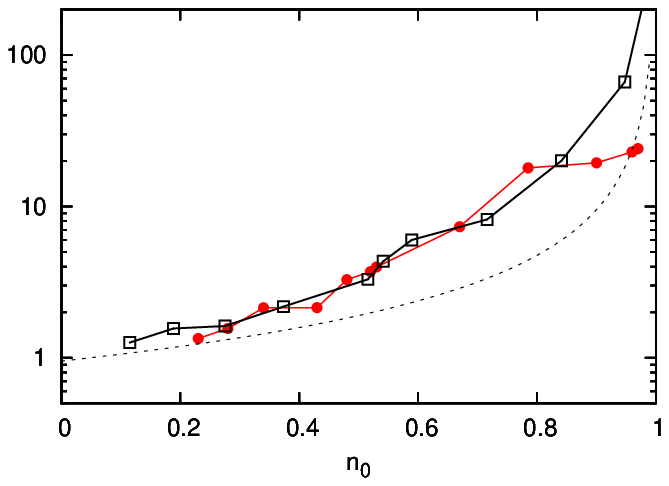}}
\put(2.0,1.5){\rotatebox{15}{quasicondensate}}
\put(6.2,2.5){BEC}
\put(1.4,4.1){\bf(e)}
\end{picture}
\vspace*{-0.3cm}
\caption{
\label{xi}
Coherence length measures. (a-d): $g^{(1)}(z,-z)$ correlation function. 
(e): Phase domain widths as a function of condensate fraction $n_0$. 
Red dots: $l_{1/2}$ values at equilibrium $t=t_{\rm max}$;
open squares: $d$, mean distance between solitons at $t=t_r$;
dashed line: stationary state prediction for $l_{1/2}$ ($l_{1/2}=l_{\phi}/\log 2=(3/2z_{TF}\log 2)(N/T_c)(\frac{1}{1-n_0})$ from \cite{Petrov00}).
}
\end{figure}

Let us now consider the thermal equilibrium state at $t=t_{\rm max}$. We observe that there are indeed no solitons present. 
The phase domain size calculated from $g^{(1)}$ is shown by solid circles Fig.~\ref{xi}(e), where we also compare to the equilibrium theoretical prediction of \cite{Petrov00} (dashed). The agreement is quite good deep in the quasicondensate regime, with a $2-3\times$ discrepancy for $0.5\lesssim n_0\lesssim0.8$. 
We note that experiments also tend to show coherence lengths longer than the theoretical estimate shown\cite{Dettmer01,Richard03}.

Having calibrated our results with the two phase fluctuating situations, a connection between them can be made. 
We observe that the soliton creation 
timescale here is very short in comparison with the thermalisation time. A hypothesis is that initial contact of coherent phase domains formed via the KZM results in a soliton, which then converts into smooth phase fluctuations only on the long thermalisation timescale. 
Such a temporal mismatch between phase and density fluctuation behaviour is also reminiscent of recent experiments in ultracold Helium\cite{TruscottUP}.
A symptom of such a process would be to see the disappearance of solitons without change in the coherence length. 
Indeed, in the quasicondensate, Figs.~\ref{xi}(a,c) show no appreciable change from $t_r$ to $t_{\rm max}$, despite the disappearance of solitons in the meantime. 
A ``soliton'' measure of the phase domain size is the mean distance $d$ between them. If phase fluctuations are ``born'' as solitons, then $d$ at earlier times should match the final phase domain width $l_{1/2}$ in equilibrium. Fig.~\ref{xi}, compares $d(t_r)$ at the end of the ramp with $l_{1/2}$ at equilibrium. The match is remarkably good, strongly backing up the above hypothesis that solitons can be considered the precursors or seeds of the equilibrium phase fluctuations when the gas is evaporatively cooled.

In 2D, where vortices take the roles of solitons in 1D, a similar sequence of defect creation and slow dissipation into long-wavelength phase fluctuations should be active when crossing the BKT transition\cite{BKT}. However, there, c-field simulations showed vortices still present in the thermal state\cite{SimulaBlakie,Weiler08}, while in 1D the solitons always eventually dissipate. 

In conclusion,
we have calculated the evaporative cooling dynamics of a 1D gas through to the quasicondensate and BEC regimes in an experimentally realistic model. 
The simulations confirm the action of the Kibble-Zurek mechanism for soliton formation, and their subsequent decay into long-wavelength phase fluctuations. 
The match between inter-soliton distance at the end of the ramp and phase coherence length at long times shows that the solitons are the early stage of the formation of 
the thermal phase fluctuations.

It is a pleasure to thank 
Wojciech \.Zurek, 
Tomasz \'Swis\l ocki,
and 
Alice Sinatra 
for valuable discussions on this topic.
This work was supported by Polish Government Research Funds: 
N N202 104136, 2009-2011 (EW,MG),  
N N202 128539, 2010-2012 (PD),
N N202 174239, 2010-2011 (KR).

\newcommand{\etal}{\textsl{et al.}}

\end{document}